\documentclass[a4paper]{jpconf}
\usepackage{graphicx}
\usepackage{mathabx}
\begin{document}
\title{Ultra high energy cosmic rays: the highest energy frontier }

\author{Jo\~ao R. T. de Mello Neto}

\address{Instituto de F\'isica, Universidade Federal do Rio de Janeiro, Ilha do Fund\~ao\\
Rio de Janeiro, RJ  Brazil}

\ead{jtmn@if.ufrj.br}

\begin{abstract}
Ultra-high energy cosmic rays (UHECRs) are the highest energy messengers of the present universe, with energies up to $10^{20}$ eV.  Studies of astrophysical particles (nuclei, electrons, neutrinos and photons) at their highest observed energies have implications for fundamental physics as well as astrophysics. The primary particles interact in the atmosphere and generate  extensive air showers. Analysis of those showers enables one not only to estimate the energy, direction and most probable mass of the primary cosmic particles, but also to obtain information about the properties of their hadronic interactions at an energy more than one order of magnitude above that accessible with the current highest energy human-made accelerator. In this contribution we will review the state-of-the-art in UHECRs detection.  We will present the leading experiments  Pierre Auger Observatory and Telescope Array  and discuss the cosmic ray energy spectrum, searches for directional anisotropy, studies of mass composition, the determination of the number of shower muons (which is sensitive to the shower hadronic interactions) and the proton-air cross section. 
\end{abstract}

\section{Introduction}
The origin and nature of the ultra high energy cosmic rays, first detected about 50 years ago \cite{Linsley:1963km}, remain unknown.  They explore the highest  energies and  kinematic regions not  directly accessible at accelerators, connecting extreme astrophysical systems with particle physics.  
For energies up to $10^{15}$ eV, cosmic rays are believed to have a galactic origin and shock acceleration in supernova remnants could be the most likely mechanism. At the highest energies, the most probable sources of  UHECRs  are extragalactic:   jets of active galactic nuclei (AGN), radio lobes, gamma-ray bursts and colliding galaxies, among others \cite{kotera}.
 In this paper we summarize\footnote{Whenever possible the results presented here are taken from the proceedings of the 34th International Cosmic Ray Conference, August, 2015, The Hague, The Netherlands.} the main experimental results from the Pierre Auger Observatory and from the Telescope Array  on measurements of UHECRs, the highest energy particles measured on Earth, with energy  E $\gtrsim 0.01$ EeV (1 EeV $ =  10^{18}$ eV).

\section{The Pierre Auger Observatory}

The Pierre Auger Observatory (Auger) ~\cite{observ}  is a hybrid detector\footnote{ It employs a hybrid technique for detection that consists in the simultaneous observation by a ground array of particle detectors and by fluorescence telescopes that are able to trace the development of the air shower in the atmosphere.} located in the Province of Mendoza, Argentina,  that combines both surface and fluorescence detectors at the same site, low energy extensions included 
\cite{amiga} \cite{heat}.   The surface detector (SD) consists of 1660 10 m$^2$ $\times$ 1.2\,m   water-Cherenkov stations  deployed over 3000
km$^2$ on a 1500 m triangular grid.   An  ``infill'' array with a 750 m grid was added to the SD with the purpose of measuring showers of  lower energy. The SD is overlooked by a fluorescence detector (FD) composed of four fluorescence stations, each one with 6 wide angle telescopes, and one additional station with 3 high-elevation telescopes also conceived to measure showers of lower energy, the High Elevation Auger Telescopes (HEAT). The surface detector   stations
sample   the electrons, photons and muons   in the shower front at ground level. The fluorescence telescopes  can record   ultraviolet 
light emitted as the shower
crosses the atmosphere, allowing  one  to observe  the longitudinal development of the air shower.
The fluorescence detector operates only on clear, moonless nights, so its duty cycle is about 13\%. On the other hand, the  surface detector array has a duty cycle close to 100\%. 

\section{The Telescope Array}
The Telescope Array experiment (TA), located in Millard Country, UT, USA, has been in operation in hybrid mode since the year 2008.  The TA has three fluorescence detector stations overlooking a surface detector array of 507 counters, each consisting of 2 layers of 3m$^2$ $\times$ 1.2 cm scintillators \cite{TA}.  The counters are spread over approximately 700 km$^2$ on a  square grid with a spacing of 1.2 km.  Two of the FD stations, Black Rock Mesa and Long Ridge, have 10 telescopes each, with 256 pixels per telescope that use a 10 MHz FADC readout system, and each station covers $108^{\circ}$ in azimuth and $3^{\circ}$ to $33^{\circ}$ degrees in elevation \cite{TA_FD}.  The Middle Drum station has 14 telescopes, with 256 pixels per telescope that use sample and hold electronics, and the station covers $112^{\circ}$ in azimuth and $3^{\circ}$ to $31^{\circ}$ in elevation \cite{TA_FD2}.   A low energy extension (TALE)  consists of additional fluorescence telescopes added near the Middle Drum site and an infill array of the same scintillation counters as used in the main array, placed at distances  1.5 to 3 km away from the Middle Drum FD.  
The end-to-end  absolute calibration of the energy is provided by a electron linear accelerator on site.  
	
\section{The energy spectra}

The all-particle energy spectrum is the most outstanding observable in cosmic ray physics, since it contains information in a combined way about the sources and about the galactic and/or intergalactic media in which the cosmic rays propagate.

The energy spectrum was measured by Auger  \cite{spectrum}, \cite{AU_Spec_ICRC} using four independent data sets:  the SD main array vertical events (zenith angle from $0^{\circ}-60^{\circ}$), threshold energy $3\times 10^{18}$ eV, 
the SD main array inclined events (zenith angle from $60^{\circ}-80^{\circ}$), threshold energy $4\times 10^{18}$ eV, the SD `infill" array, threshold energy $3\times 10^{17}$ eV  and the hybrid sample,  (zenith angle from $0^{\circ}-60^{\circ}$), threshold energy $ 10^{18}$ eV.
The absolute calibration of the SD is inferred from a high-quality subset of hybrid events used to calibrate the SD energy estimators using the calorimetric energies measured with the FD.  The SD  shares the uncertainty of the FD energy scale of 14\%. 
 
\begin{figure}[h]
\begin{minipage}{18pc}
\includegraphics[width=20pc]{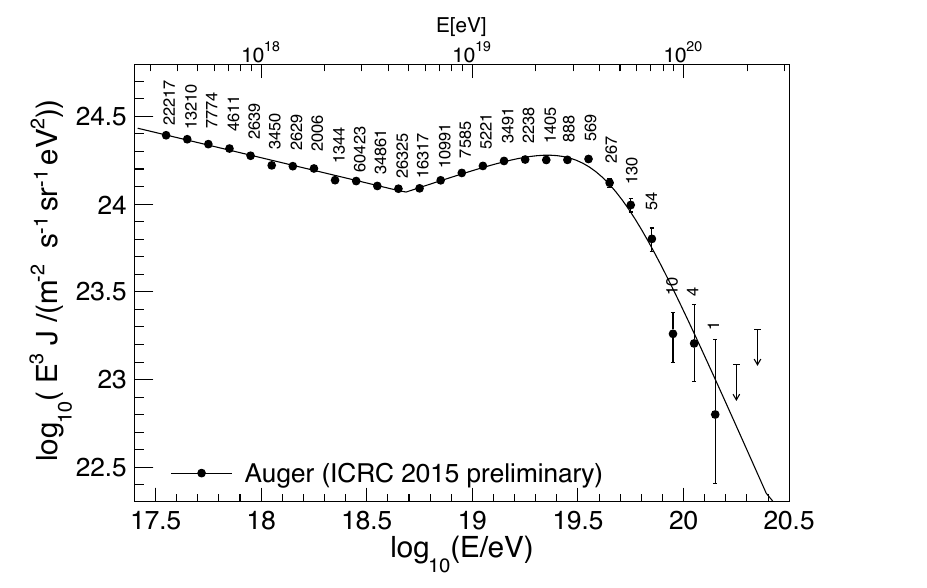}
\caption{\label{AU_spec} The combined energy spectrum fitted with a flux model.  Only statistical uncertainties are shown.  The systematic uncertainty on the energy scale is 14\%. The number of events is given above the points and the upper limits correspond to the 84\% C.L. \cite{AU_Spec_ICRC}.}
\end{minipage}\hspace{2pc}%
\begin{minipage}{18pc}
\includegraphics[width=18pc]{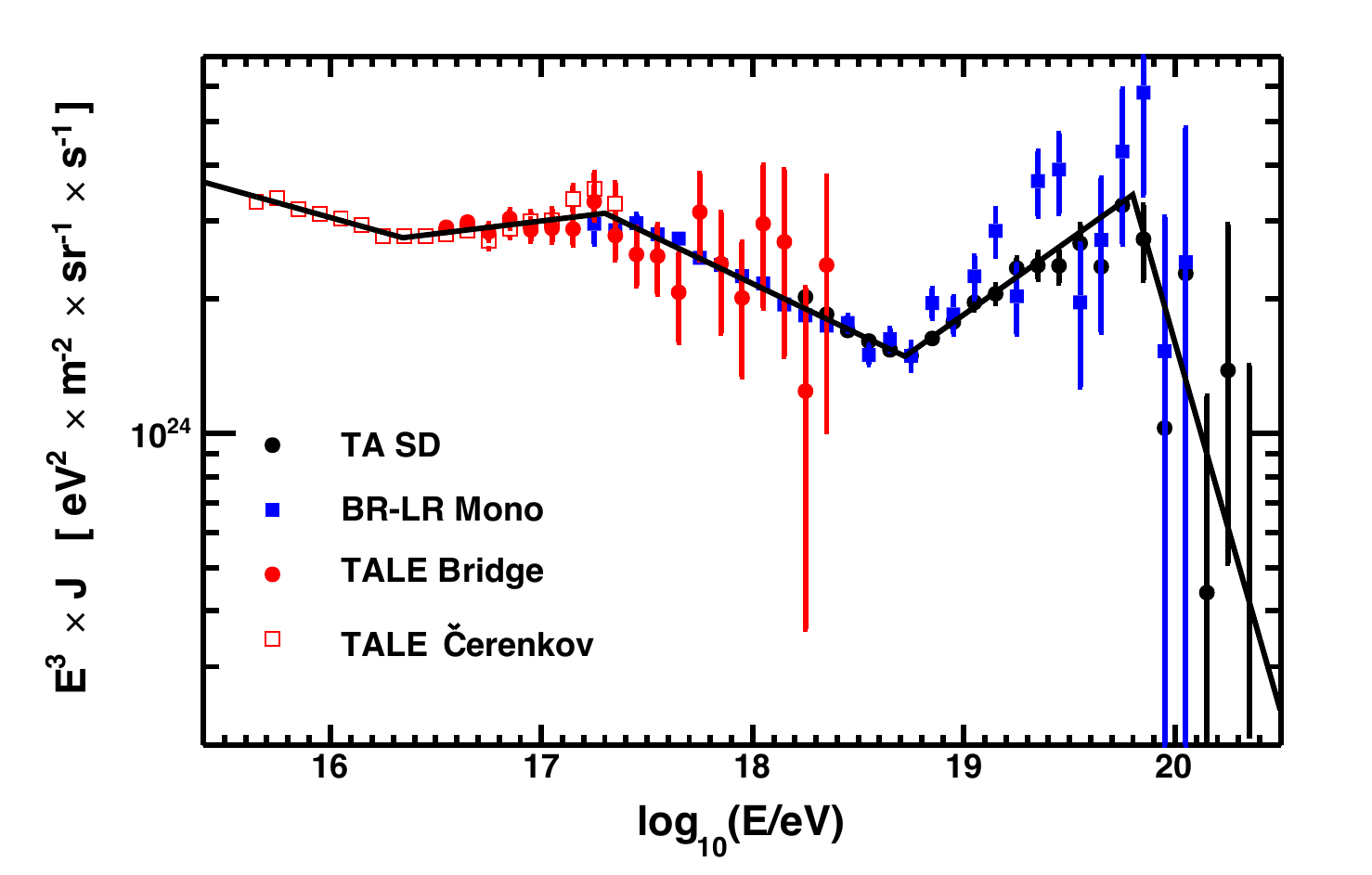}
\caption{\label{TA_spec} The combined energy spectrum measured by TA contained TA SD (E$\, >$$10^{18.2}$\,eV), TA FD monocular (from Black Rock Mesa and Long Ridge, E$ \, > 10^{17.2}$ eV),  TALE FD ( $10^{16.5} < $ E  $ < 10^{18.4}$\, eV) and TALE FD reconstructed using only the Cherenkov light  ( $10^{15.6} < $ E $ < 10^{17.4}$ eV) \cite{TA_Spec_ICRC}.}
\end{minipage} 
\end{figure}

The final step in measuring the energy spectrum is a precise determination of the exposure for the observations. Above the energy for the full detector efficiency, the calculation of the SD exposure is based solely on the determination of the geometrical aperture of the array for the corresponding zenith-angle interval and of the observation time. The exposure of the hybrid mode of Auger has been calculated using a time-dependent Monte Carlo simulation.  The result is an exposure growing with shower energy above the threshold energy of $10^{18}$ eV.  A correction is applied to the measured flux to account for the effect of the finite resolution in the energy determination, responsible for bin-to-bin migration. 
All four spectra are in agreement within uncertainties.  In Fig. \ref{AU_spec} they are combined using a method that takes into account the systematic uncertainties of the individual measurements.

The TA SD spectrum was calculated using data from 2008/05/11 to 2015/05/11 \cite{TA_Spec_ICRC}.  First the event geometry and the lateral distribution are reconstructed.  Next, a Monte Carlo simulation using CORSIKA, QGSJET-II.3 and GEANT4 is done and the response of each TA SD unit is obtained.  The simulated events are reconstructed using the same quality cuts and procedures as the data.  The TA SD Monte Carlo (MC) uses proton composition and the spectrum measured by HIRES and TA \cite{TA_Spec}.  A energy estimation table derived from the MC is used to reconstruct energies in both data and Monte Carlo.  The energy scale is corrected to the TA FD using events detected by both the FD and SD.   Fig. \ref{TA_spec}  shows the combined spectrum measured by TA contained TA SD, TA FD monocular (from Black Rock Mesa and Long Ridge),  TALE FD and TALE FD reconstructed using only the Cherenkov light.

The energy spectra of both observatories are compatible within the systematic uncertainties.  Both observe the ``ankle'' at E $\approx 4.8$  EeV (Auger)  and E $\approx 5.2 $  EeV (TA).  The data also show  that the flux-suppression in the Auger data starts at  lower energies and also falls off more strongly than in TA data:  E$_{1/2}$, the energy at which the differential flux falls to one-half of the value of the power-law extrapolation from the intermediate region, is about 25 EeV for Auger and 60 EeV for TA. 

The suppression at the highest energies, measured with unprecedented statistical significance by both experiments, 
is consistent with the expectations from the so called `GZK suppression", understood as the attenuation of extragalactic protons by photo-pion production off CMB photons or as the suppression of  nuclei by photo-desintegration. However, it must  be noted that the interpretation of the flux suppression in terms of interactions with the CMB does not exclude	additional contributions related to the acceleration mechanism, such as a change in the injection spectrum at the source or the maximum energy of the accelerators. 

\section{Mass composition}
The measurement of the mass composition of UHECRs is essential to the solution of the problem of their origin, since the mass and charge $Z$  distributions can give powerful constraints on their acceleration mechanisms and propagation.
For UHECRs the main observable sensitive to composition is the $\langle X_{\textrm{max}} \rangle$, the average value of the atmospheric depth (measured in g/cm$^2$) where the shower development reaches its maximum. Proton showers have   $\langle X_{\textrm{max}} \rangle$ about 100 g/cm$^2$ deeper in the atmosphere than iron showers. In a similar way, the fluctuation of the values of $X_{\textrm {max}}$ around the mean depth, $\textrm{RMS}(X_{\textrm {max}})$, provides another sensitive observable: iron showers fluctuate about 40 g/cm$^2$ less than proton showers.  Those estimates are obtained from transport codes that simulate  shower development given a model for hadronic interactions.

\begin{figure}[h]
\begin{minipage}{18pc}
\includegraphics[width=18pc]{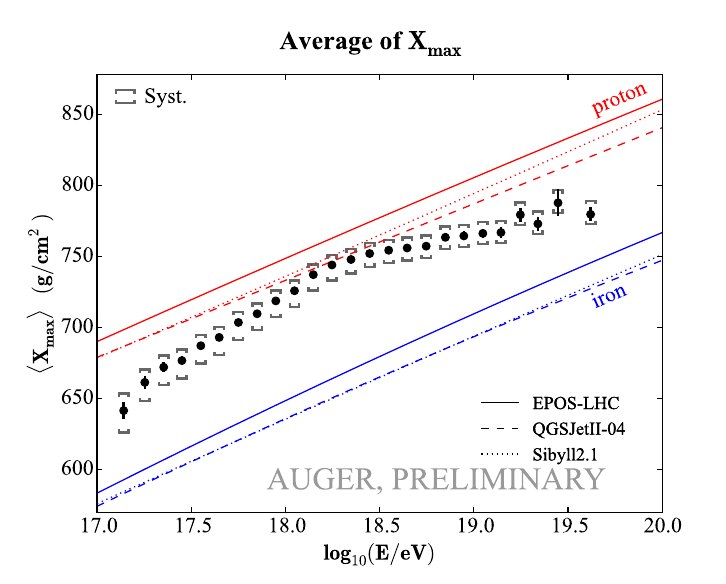}
\caption{\label{xmaxAU} $\langle X_{\textrm {max}} \rangle$ as a function of energy  as measured by the Pierre Auger Observatory  (combining HeCo and standard datasets) as a function of energy compared to air-shower simulations for proton and iron primaries \cite{AU_Xmax_ICRC}. }
\end{minipage}\hspace{2pc}%
\begin{minipage}{18pc}
\includegraphics[width=18pc]{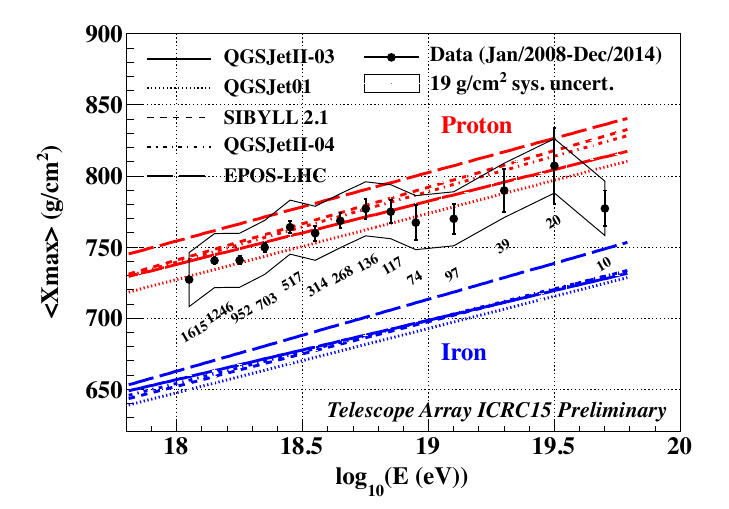}
\caption{\label{xmaxTA}  $\langle X_{\textrm {max}} \rangle$ as a function of energy  from data (black points) from the Telescope Array compared to air-shower simulations for  proton and iron and five hadronic interaction models. The box region shows the systematic uncertainty on $\mathrm{X_{max}}$,  19 g/cm$^2$, for the monocular analysis. \cite{TA_Xmax_ICRC}}
\end{minipage} 
\end{figure}

The measurements by Auger cover nearly three decades of energy with the addition of the HEAT enhancements. The HEAT telescopes cover from $30^{\circ}$ to $60^{\circ}$ in elevation and are located near to one of the standard FD sites (Coihueco).  The combination of the HEAT and Coihueco FD detectors (HeCo dataset) covers from $2^{\circ}$ to $60^{\circ}$ in elevation, making it possible to reach showers with $10^{17}$ eV.   Fig. \ref{xmaxAU} shows that between $10^{17.0}$ and $10^{18.3}$ eV $\langle X_{\textrm {max}} \rangle$ increases by $\approx$ 85 g cm$^{-2}$ per decade of energy.  One expects about 60 g 
cm$^{-2}$ per decade for a constant composition, so this means that the mean primary mass is getting lighter.  Around $\approx 10^{18.3}$ eV  the rate of change of $\langle X_{\textrm {max}} \rangle$ becomes smaller, about 26  g cm$^{-2}$ per decade, indicating a transition to a heavier composition.  The fluctuations of $\langle X_{\textrm {max}} \rangle$ start to decrease at about  the same energy $\approx 10^{18.3}$ eV \cite{AU_Xmax_ICRC}.  

TA has measured  $X_{\textrm {max}} $   in several ways, which is very convenient for systematic errors checks.                                                 Here we report   on                                                                                                                                                                                                                                                                                                                                                         the latest TA results obtained with  data collected  at the newly constructed FD stations using a monocular analysis, which is an analysis mode to reconstruct an EAS to obtain properties of the primary particles using the measured shower image by one FD station \cite{TA_Xmax_ICRC}. Fig. \ref{xmaxTA} shows the comparison between observed $X_{\textrm {max}} $  and Monte Carlo simulations.  The systematic uncertainty on $\langle X_{\textrm {max}} \rangle$ is 19 g cm$^{-2}$ for the monocular analysis indicated in the box region shown in the figure.  TA measurements are consistent with proton-dominated composition at this energy range, as reported in previous TA publications. 

The difference between the Auger and TA interpretation of the $X_{\textrm {max}} $  was addressed in a joint working group from both collaborations.  Auger analysis is based on unbiased cuts, while TA folds the data with detector effects. When TA reconstruct simulated events compatible with the $X_{\textrm {max}} $ distribution from Auger and compare this simulation with TA data, there is a very good agreement.  So,  after accounting for the different resolutions, acceptances and analysis strategies of the two experiments, the two results are found to be in good agreement within systematic uncertainties \cite{WG_Xmax_ICRC}.

\section{Anisotropies}

\begin{figure}[h] 
\begin{minipage}{18pc}
\includegraphics[width=18pc]{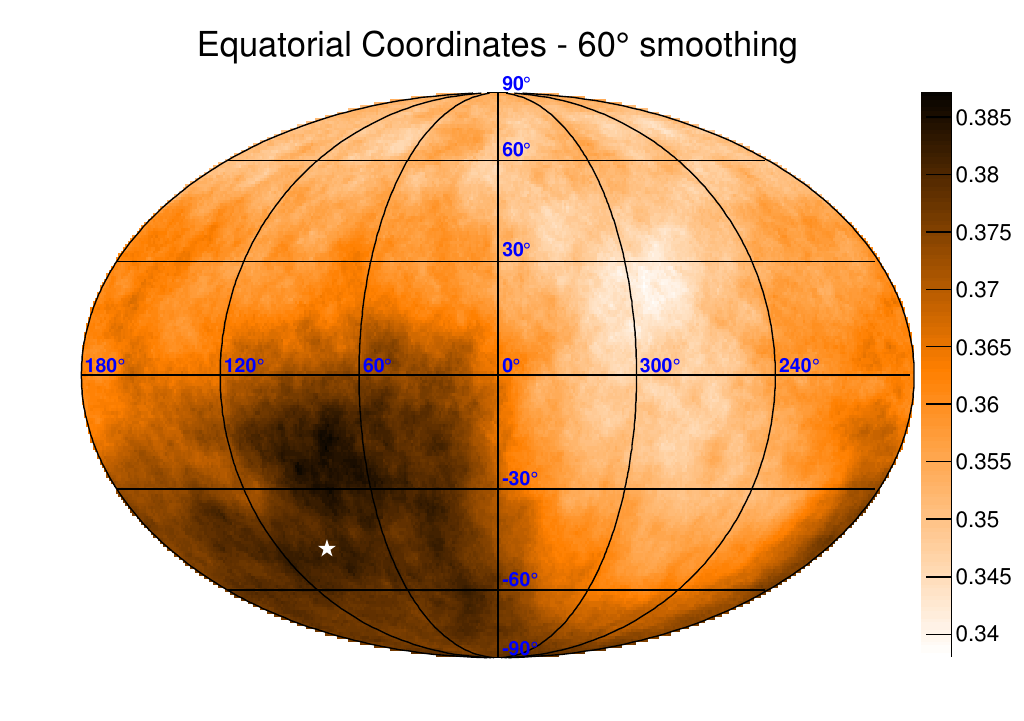}
\caption{\label{skymap} Sky map in equatorial coordinates of the average flux measured by Auger and TA smoothed out at a $60^{\circ}$ angular scale (to exhibit the dipole structure)  above $10^{19}$  eV in km$^{-2}$ yr$^{-1}$  sr$^{-1}$ units. The direction of the reconstructed dipole is shown as the white star \cite{fullSky_ICRC}.}
\end{minipage}\hspace{2pc}%
\begin{minipage}{18pc}
\includegraphics[width=18pc]{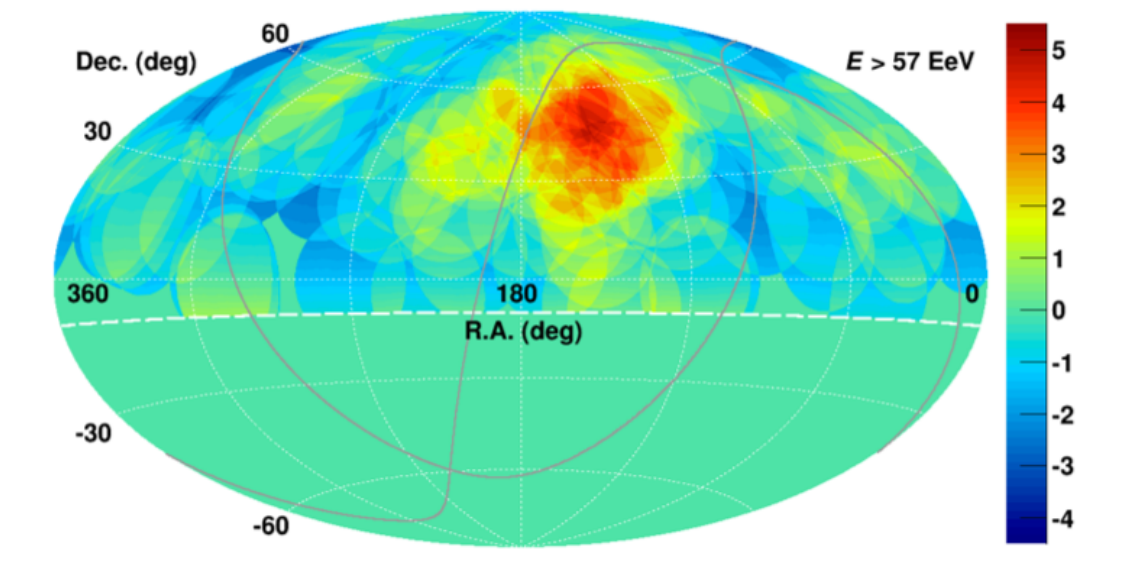}
\caption{\label{hotspot}    Significance map for the 7-year observation by TA  using a $20^{\circ}$  oversampling radius. The maximum significance is 5.1$\sigma$. The map uses Aitoff projection in equatorial coordinates. The solid curves indicate the galactic plane and supergalactic plane. \cite{TAhotspot_ICRC}}
\end{minipage} 
\end{figure}

Another very  important observable that sheds light on the nature and origin of UHECRs is the distribution of their arrival directions over the sky.  Their arrival directions are basically free from systematic errors, in contrast with energies or primary mass.  Although the sources of UHECRs are yet to be discovered, their large-scale distribution is expected to follow the local distribution of matter in the universe at some level.  Dipoles, quadrupoles and higher order multipoles of the distribution in the sky could be present due to diffusive propagation of UHECRs, excesses in the super-Galactic plane, and other possible features of the source distributions.   With combined data from the  Auger and TA collaborations, multipole coefficients of the UHECR flux  from a full-sky coverage were measured  \cite{fullSky} and updated recently \cite{fullSky_ICRC}.  The key issue of this work is that an unambiguous measurement of dipole and quadrupole moments as well as of the full set of spherical harmonic coefficients requires full-sky coverage.  To this end, a joint analysis using data recorded at  Auger and the TA  has been performed for energies above E  = $10^{19}$  eV in terms of the  energy scale used by TA. The Auger  energy threshold was taken as the value that guarantees equal fluxes for both experiments.  The band of declinations between $-15^{\circ}$ and $45^{\circ}$  is accessible to the fields of view of both experiments.  This overlapping region was used to estimate the multipole coefficients of the flux expansion through an iteration method   by adjusting a parameter that re-weights the directional Auger exposure.  From TA were left  2,560 events (1,703 in the common declination band) above  E $ = 10^{19} $ eV and 16,835 (5,885 in the common band) above $8.8\times 10^{18}$ eV from Auger.  The dipole amplitude is observed to be $(6.5 \pm 1.9)$\% with a chance probability of $ 5\times 10^{-3}$, pointing to $(93^{\circ} \pm 24^{\circ}$) in right ascension and $(-46^{\circ} \pm 18^{\circ})$ in declination, as shown in Fig. \ref{skymap} \cite{fullSky_ICRC}. The results are in agreement with the ones published by Auger \cite{dipole} without assumptions on the underlying flux of UHECRs.

Fig. \ref{hotspot} displays the significance map showing the TA ``hotspot'', an excess of events with  E $ > 57$~EeV and zenith angle $\theta < 55^{\circ}$  centered near the Ursa Major cluster (R.A. $(\alpha) = 148.5^{\circ}$, Dec. $(\delta) = 44.6^{\circ}$).   The total number of events is 109, taken from 2008 May 11 to 2015 May 11.  The chance probability of such a hotspot appearing by chance anywhere in the sky is $ 3.7\times 10^{-4}$, equivalent to a one-sided probability of 3.4$\sigma$ \cite{TAhotspot_ICRC}.  A few powerful  celestial objects  around the hotspot, such as  blazars Mrk 421, Mrk 180 and starburst galaxy M82 have been suggested as the candidates of its origin \cite{hotSource1, hotSource2}.

\section{Hadronic interactions}

\begin{figure}[h]
\begin{minipage}{18pc}
\includegraphics[width=18pc]{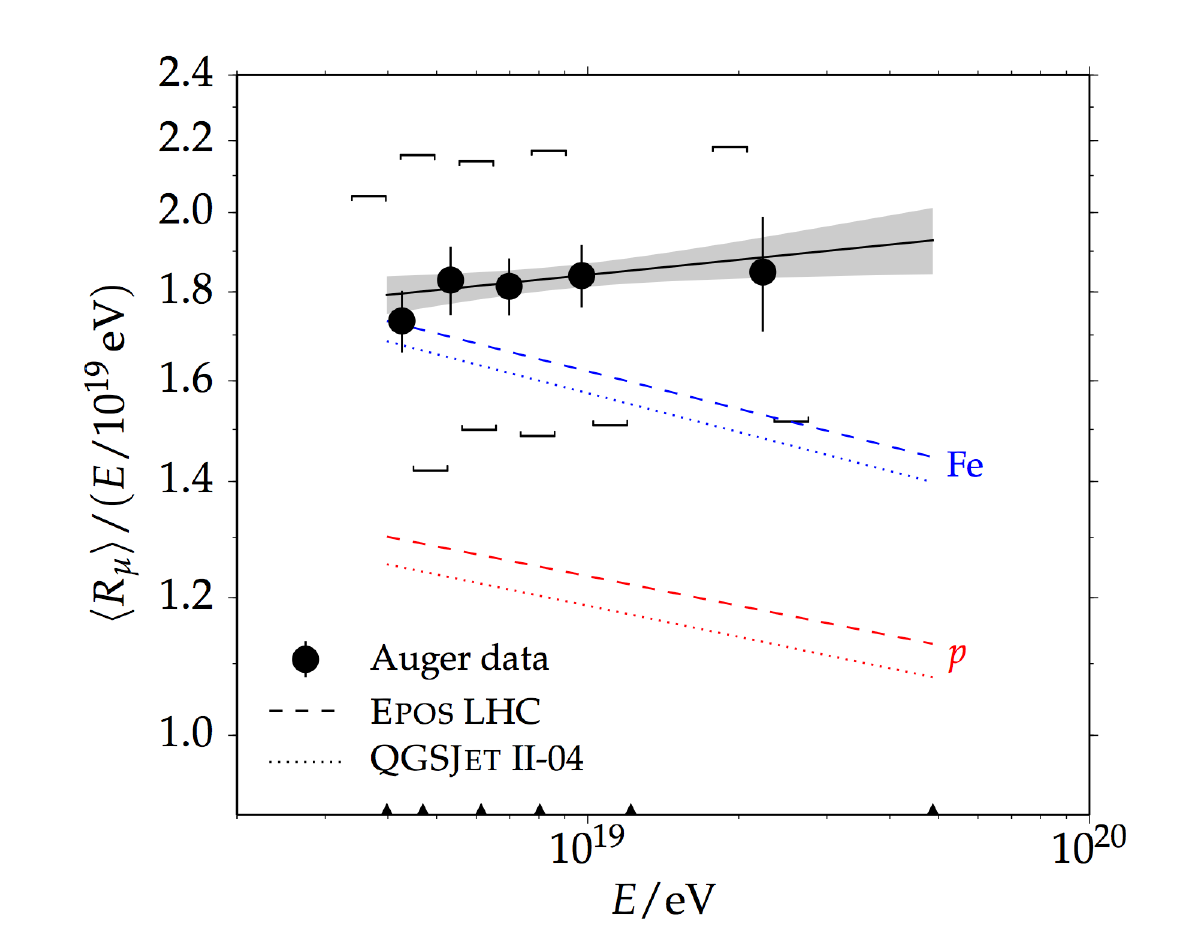}
\caption{\label{muons} $\langle R_{\mu} \rangle$ as a function of primary energy, compared to shower simulations of two hadronic models for proton and iron \cite{AUmu_ICRC}}
\end{minipage}\hspace{2pc}%
\begin{minipage}{18pc}
\includegraphics[width=18pc]{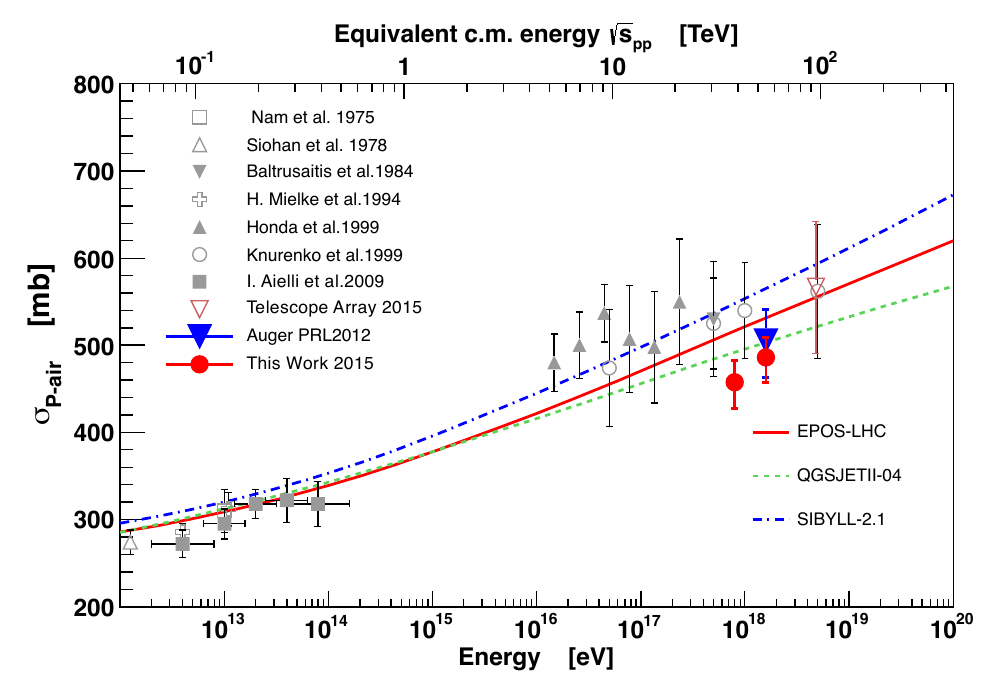}
\caption{\label{xsec} The  $\sigma_p$-air measurement compared to previous data and model predictions. \cite{AUxsec_ICRC}}
\end{minipage} 
\end{figure}

Apart from  $X_{\textrm {max}}$, another estimator of the composition of the primary particles is the muon density at 	ground level.  Techniques to reconstruct inclined showers \cite{AUinclined} have been used by Auger to extract the muon content of air showers.  In those events the dominant particles at ground are muons since the electrons and photons were absorbed by the atmosphere.
 The observable $N_{19}$  gives the number of muons per unit area relative to the reference density, which is obtained from proton simulations at $10^{19}$  eV using the QGSJETII-03 model for hadronic interactions. 
 For inclined showers, the energy of the primary is obtained by calibrating $N_{19}$ with the calorimetric energy $E_{FD}$ from high-quality events measured simultaneously with the fluorescence detector and the surface detector for inclined showers. The total number of muons $N_{\mu}^{\textrm{est}}$  can be obtained as the product of $N_{19}$  times the surface integral of the reference functions for the number density of muons.  Then a ratio estimator  $\langle R_{\mu} \rangle $ that gives the relative number of muons with respect of the reference density is calculated.  In Fig. \ref{muons} the averages of $R_{\mu}$, divided by the energy, are plotted for five energy bins and compared to simulations for protons and irons showers for the QGSJETII-04 and EPOS-LHC hadronic models at $\langle \theta_{\textrm{data}} \rangle = 67^{\circ}$.  The predictions for proton and iron are well separated, showing that $R_{\mu}$ is a good estimator.  The measured muon number is higher  than that expected in pure iron showers.  This is not in agreement with studies based on  $X_{\textrm {max}} $ \cite{antoine}  that point to an average logarithmic mass $\langle \ln A \rangle$ being between proton and iron in this energy range \cite{AUmu_ICRC}.   These results, and others based  on different approaches,  point to the fact that simulations of proton primaries and iron primaries underestimate the muon fraction at ground level as measured by Auger.

An update of the analysis of the proton-air cross-section based on the shape of the distribution of $X_{\textrm{max}}$ published  in \cite{xsec} was presented in \cite{AUxsec_ICRC}.  The analysis is based on the fact that the tail of the $X_{\textrm{max}}$ distribution is sensitive to $\sigma_{\textrm{p-air}}$. 
The cross section is  related to the exponential distribution of the depth of the first interaction $X_1$ which cannot be measured.  But the strong correlation between $X_1$ and $X_{\textrm{max}}$ makes the distribution of the latter sensitive to the proton-air cross-section and the tail of the distribution maximizes the proton content, since it is the most penetrating nucleus.   A slope obtained from a fit to the exponential tail of the $X_{max}$ distribution can be used as an estimator for $\sigma_{p-air}$ through Monte Carlo simulations: the cross-section is rescaled consistently to reproduce the value of the measurement. The lack of detailed knowledge of the mass composition at these energies turns out to be the main difficulty for this analysis, since one cannot exclude contamination by photons and helium primaries, for instance. This translates into the main contribution to the systematic uncertainty of this measurement. 
The available data sample is divided into two energy intervals, one ranging from $10^{17.8}$ to $10^{18}$ eV and the other from $10^{18}$ to $10^{18.5}$  eV with 18090 and 21270 events and the measured cross-sections are $457.5 \pm 17.8 ({\rm stat})  ^{+19}_{-25} ({\rm sys})$ mb and $485.8 \pm 15.8 ({\rm stat})  ^{+19}_{-25} ({\rm sys})$ mb respectively. 
While the composition of primary cosmic rays in the above energy ranges is compatible with being dominated by protons, a contamination with Helium cannot be excluded.  The quoted systematic uncertainties take into account, among many other effects, an impact of 25\% Helium in the data sample. 
Fig. \ref{xsec} displays the $\sigma_p$-air measurement compared to previous data and model predictions.  The data are consistent with a rising cross section with energy, however, the statistical precision is not yet sufficient to make a statement on the functional form \cite{AUxsec_ICRC}. 

 \section{Conclusions and outlook}
 
 The Pierre Auger Collaboration plans to complement the water-Cherenkov detectors of the surface array with scintillators to determine the muonic shower component.  This will extend the composition sensitivity of the experiment  into the flux suppression region.  It will also allow the estimation of the primary mass of the highest energy cosmic rays on a shower-by-shower basis.  The measurement of the mass composition, the search for light primaries at the highest energies, the study of composition-selected anisotropy  and the search for new phenomena including unexpected changes of hadronic interactions are the main objectives of the upgrade.  The upgrade is named AugerPrime and it is proposed to run from 2018 to 2024 \cite{AugerUpgrade}.

The Telescope Array collaboration plans an extension (called TAx4)   to quadruple the area of the TA SD array to approximately 3,000 km$^2$, by adding 500 surface detectors with 2.08 km spacing. Two FD stations will be constructed viewing the new SD array.  The upgrade was approved in Japan in April, 2015 and a proposal will be submitted in the US in October 2015. 
The main objectives are the confirmation of the hotspot at a post-trial significance greater than $5\sigma$, search for its
origin  and  to enhance TA's cosmic ray energy spectrum measurements and composition studies at the highest energies \cite{TAupgrade}. 

Together with other astroparticle projects, the Pierre Auger Observatory and the Telescope Array will continue to provide first quality data that surely will have a strong impact in the field.

\end{document}